\documentclass[prb,
twocolumn,
superscriptaddress,showpacs,amsmath,amssymb,longbibliography]{revtex4-1}

\usepackage{xcolor}
\usepackage{graphicx}
\usepackage{hyperref}

\begin{document}

\title{Vortices in polar and $\beta$ phases of $^3$He}

\author{G.E.~Volovik}
\affiliation{Low Temperature Laboratory, Aalto University,  P.O. Box 15100, FI-00076 Aalto, Finland}
\affiliation{Landau Institute for Theoretical Physics, acad. Semyonov av., 1a, 142432,
Chernogolovka, Russia}

\date{\today}

\begin{abstract}
Recently new topological  phase of superfluid $^3$He has been discovered -- the $\beta$ phase, which is obtained by strong polarization of the nematic polar phase. We consider half-quantum vortices, which are formed in rotating cryostat, and discuss the evolution of the vortex lattice in the process of the transition from the polar phase to the $\beta$-phase via the spin-polarized polar phase.
\end{abstract}
\pacs{}

\maketitle 
 
\section{Introduction}

At the moment the most interesting topics in the condensed matter physics are related to topological materials: topological insulators, topological superconductors, Dirac and Weyl topological semimetals, etc. Superfluid phases of liquid $^3$He\cite{VW,Volovik2003,Mizushima2016} are the best representatives of the topological matter. Each phase has its unique topological  property. 

Recently, the new phases has been discovered, which are also unique: the polar phase\cite{Dmitriev2016,Halperin2018} and the $\beta$-phase.\cite{Dmitriev2021} 
These superfluid phases were obtained by confinement of liquid $^3$He in nematic aerogels with nearly parallel strands.
The polar phase  has Dirac nodal line in the fermionic spectrum in bulk liquid and the flat band (drumhead states)  on the surface,\cite{Eltsov1908,Autti2020}  which is similar to that in the semimetals with nodal lines.\cite{Heikkila2011,Kopnin2011}

 In the $\beta$ phase the superfluid pairing takes place only for single spin projection. Thus the spin degeneracy of the flat band is lifted, and the surface contains the non-degenerate Majorana fermions. Also, similar to the polar phase the $\beta$ phase in aerogel is robust to disorder owing to the extension of the Anderson theorem to superconductors with columnar defects.\cite{Fomin2018,Eltsov1908,Ikeda2020} 
 
 Here we consider the vortex states, which appear in the rotating polar and $\beta$ phases, and  their evolution in the process of transformation of the polar phase to the $\beta$ phase.

\section{Single-quantum vortex and half-quantum vortices}

The spin-triplet $p$-wave order parameter, $A_{\alpha i} \equiv {\bf A}_i$, in the polar phase, in the spin-polarized polar phase, and in the $\beta$-phase has the following form:
\begin{eqnarray}
{\bf A}_i= \hat z_i 
\left(\Delta_{\uparrow \uparrow}(\hat{\bf x} + i \hat{\bf y})e^{i\Phi_\uparrow} +
 \Delta_{\downarrow \downarrow}(\hat{\bf x} - i  \hat{\bf y})e^{i\Phi_\downarrow}\right) \,
\label{order_parameter}
\end{eqnarray}
where $\Delta_{\uparrow \uparrow}= \Delta_{\downarrow \downarrow}$ in the polar phase,
$\Delta_{\downarrow \downarrow}<\Delta_{\uparrow \uparrow}$ in the spin-polarized polar phase in the magnetic field along $z$-axis, and $\Delta_{\downarrow \downarrow}=0$ in the $\beta$-phase.

In rotating vessel there is the competition between the energy of two half-quantum vortices (HQVs) and the energy  of a single-quantum vortiex (SQV).\cite{Autti2016,Makinen2019}
The SQV is the phase vortex with $\Phi_\uparrow=\Phi_\downarrow =\phi$, where  $\phi$ is the azimuthal angle in cylindrical coordinates:
\begin{eqnarray}
{\bf A}_i=e^{i\phi}  \hat z_i 
\left(\Delta_{\uparrow \uparrow}(\hat{\bf x} + i \hat{\bf y}) + \Delta_{\downarrow \downarrow}(\hat{\bf x} - i  \hat{\bf y})\right) \,.
\label{SQV}
\end{eqnarray}

In the HQVs either $\Phi_\uparrow =\phi$ and $\Phi_\downarrow=0$, or $\Phi_\downarrow =\phi$ and $\Phi_\uparrow=0$:
\begin{eqnarray}
{\bf A}_i= \hat z_i
\left(\Delta_{\uparrow \uparrow}(\hat{\bf x} + i \hat{\bf y})e^{i\phi} + \Delta_{\downarrow \downarrow}(\hat{\bf x} - i  \hat{\bf y}) \right) =
\nonumber
\\ = e^{i\frac{\phi}{2}} \hat z_i
\left(\Delta_{\uparrow \uparrow}(\hat{\bf x} + i \hat{\bf y}) e^{i\frac{\phi}{2}} + \Delta_{\downarrow \downarrow}(\hat{\bf x} - i  \hat{\bf y}) e^{-i\frac{\phi}{2}}  \right) \,,
\label{up}
\end{eqnarray}
\begin{eqnarray}
{\bf A}_i= \hat z_i
\left(\Delta_{\uparrow \uparrow}(\hat{\bf x} + i \hat{\bf y}) + \Delta_{\downarrow \downarrow}(\hat{\bf x} - i  \hat{\bf y})e^{i\phi} \right) =
\nonumber
\\ =  e^{i\frac{\phi}{2}} \hat z_i
\left(\Delta_{\uparrow \uparrow}(\hat{\bf x} + i \hat{\bf y}) e^{-i\frac{\phi}{2}}  + \Delta_{\downarrow \downarrow}(\hat{\bf x} - i  \hat{\bf y})  e^{i\frac{\phi}{2}}  \right).
\label{down}
\end{eqnarray}
Eq.(\ref{up}) describes the half-quantum vortex, which represents the single-quantum vortex in the spin-up component, and
Eq.(\ref{down}) describes another half-quantum vortex, which represents the single-quantum vortex in the spin-down component, see also Ref. \cite{Sauls2021}.
When two half-quantum vortices are combined, they  form  the single-quantum vortex with both spin components in Eq.(\ref{SQV}).

The energy of superflow in the cryostat  rotating  with angular velocity $\boldsymbol{\Omega}\parallel \hat{\bf z}$ is:
\begin{eqnarray}
F=\frac{1}{2} \rho_{s\uparrow}({\bf v}_{s\uparrow} -  \boldsymbol{\Omega} \times {\bf r}) ^2 +
\frac{1}{2} \rho_{s\downarrow}({\bf v}_{s\downarrow} -  \boldsymbol{\Omega} \times {\bf r}) ^2 +
\label{FlowEnergy}
\\
+\rho_{\uparrow\downarrow}({\bf v}_{s\uparrow} -  \boldsymbol{\Omega} \times {\bf r}) ({\bf v}_{s\downarrow} -  \boldsymbol{\Omega} \times {\bf r}) \,,
\label{ABterm}
\end{eqnarray}
where ${\bf v}_{s\uparrow} = (\hbar/2m)\nabla \Phi_{\uparrow}$ and ${\bf v}_{s\downarrow} = (\hbar/2m)\nabla \Phi_{\downarrow}$.
Here Eq.(\ref{ABterm}) represents the Andreev-Bashkin term \cite{AndreevBashkin1976} which mixes spin components
and stabilizes the HQVs \cite{Leggett2009}.  In the London limit, the energy of two separated HQVs is  proportional to  $(\rho_{s\uparrow}  + \rho_{s\downarrow})$, while the energy of SQV is  proportional to 
$(\rho_{s\uparrow}  + \rho_{s\downarrow} + 2\rho_{\uparrow\downarrow})$, and for $\rho_{\uparrow\downarrow}>0$ the SQV splits into two HQVs.

\section{Vortex lattices: from polar phase to $\beta$-phase}

\begin{figure}[h]
 \includegraphics[width=0.7\columnwidth]{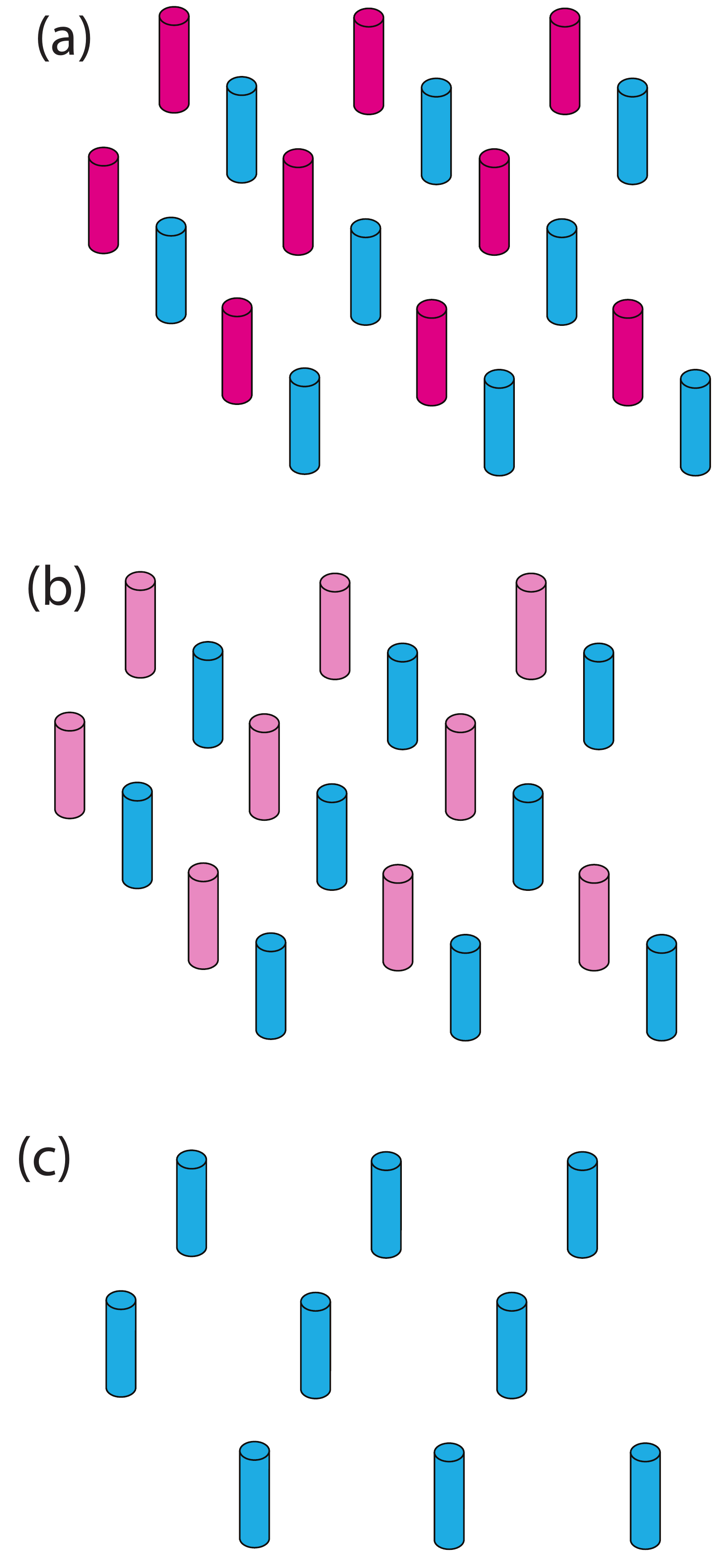}
 \caption{
Illustration of the vortex lattices in the (a) polar phase, (b) spin-polarized polar phase and (c) $\beta$-phase in rotating cryostat.
 \\
(a) The vortex lattice in zero magnetic field, where $\Delta_{\downarrow \downarrow}=\Delta_{\uparrow \uparrow}$ and $\rho_{s\downarrow}=\rho_{s\uparrow}$. The elementary cell of the lattice contains  two types of  half-quantum vortices. These are: the HQV, which can be represented  as single-quantum vortex in the spin-up  component in Eq.(\ref{up}) ({\it blue}), and  the HQV, which can be represented  as the single-quantum vortex  in the spin-down component  in Eq.(\ref{down}) ({\it red}).
The densities of vortices are in Eq.(\ref{VortexDensity}).
\\
(b) The same in the spin-polarized polar phase in non-zero magnetic field ${\bf H} \parallel \hat{\bf z}$, where the amplitude of the spin-down component $\Delta_{\downarrow \downarrow}<\Delta_{\uparrow \uparrow}$ and its superfluid density 
$\rho_{s\downarrow}<\rho_{s\uparrow}$. The densities of vortices are the same as in Eq.(\ref{VortexDensity}), but the intensity of the down-spin vortices ({\it light red}) decreases, and they finally disappear at the transition to the $\beta$-phase.
\\
(c) Vortices in the $\beta$-phase, where $\Delta_{\downarrow \downarrow}=0$ and  $\rho_{s\downarrow}=\rho_{\uparrow\downarrow}=0$. These are the single-quantum vortices in the spin-up component in Eq.(\ref{beta_vortex}). Their density $n_{\uparrow}$  is the same as in Eq.(\ref{VortexDensity}). 
\\
The area of the elementary cells is the same in all three configurations (if there is no spontaneous period doubling during the process).
 }
 \label{vortices}
\end{figure}

For $\rho_{\uparrow\downarrow}>0$ the SQV has higher energy than two HQVs. 
That is why in the polar phase and in the spin-polarized polar phase the vortex lattice splits in two sublattices of spin-up and spin-down vortices in Eqs. (\ref{up}) and (\ref{down}) correspondingly, see Fig.\ref{vortices}.
The densities of spin-up and spin-down vortices in the equilibrium vortex state  in the rotating cryostat is:
\begin{eqnarray}
n_{\uparrow} = n_{\downarrow} = \frac{2m\Omega}{\pi \hbar}\,.
\label{VortexDensity}
\end{eqnarray}

Fig. \ref{vortices}(a) demonstrates the lattice of half-quantum vortices in the polar phase in zero magnetic field, where 
$\Delta_{\downarrow \downarrow}=\Delta_{\uparrow \uparrow}$ and $\rho_{s\downarrow}=\rho_{s\uparrow}$. The elementary cell of the lattice contains two vortices: the spin-up and spin-down vortices, which have  the same energy.  The densities of vortices are in Eq.(\ref{VortexDensity}).
The spin-up and spin down vortices have opposite circulations of spin current, and thus there is no global spin current: their spin currents compensate each other.

When the polar phase is spin-polarized by the  non-zero magnetic field ${\bf H} \parallel \hat{\bf z}$, the balance between spin-up and spin-down vortices is violated. The lattice as before contains two sublattices of spin-up and spin-down voritces, see  Fig. \ref{vortices}(b). But vortices in  the spin-down component have smaller amplitude, $\Delta_{\downarrow \downarrow}<\Delta_{\uparrow \uparrow}$, smaller superfluid density, $\rho_{s\downarrow}<\rho_{s\uparrow}$, and thus the smaller energy. Also, the spin currents of spin-up and spin-down vortices are not compensated, and there is the global spin current, which is proportional to $\boldsymbol{\Omega} \times {\bf r}$. This is similar to the mechanism of the formation of the global spin currents discussed in Refs.\cite{Brauner-Moroz2019,Volovik2020}.

With increasing field the spin-down vortices are  gradually evaporated  and  finally the spin-down lattice fades away at the transition to the $\beta$-phase in Fig. \ref{vortices}(c), where $\Delta_{\downarrow \downarrow}=\rho_{s\downarrow}=\rho_{\uparrow\downarrow}=0$ and only the vortices in the spin-up component remain:
\begin{eqnarray}
{\bf A}_i=e^{i\phi} 
\Delta_{\uparrow \uparrow}\hat z_i(\hat{\bf x} + i \hat{\bf y}) \,.
\label{beta_vortex}
\end{eqnarray}
The lattice of these single-quantum vortices in the spin-up component has the same vortex density $n_{\uparrow}$ in Eq.(\ref{VortexDensity}) as in the polar phase.

\section{Vortex pinning and metastable configurations}

In the above consideration we ignored the pinning of vortices by aerogel strands. However, the pinning is rather strong.\cite{Autti2016,Makinen2019}
It leads in particular  to the formation of different vortex glasses,\cite{Eltsov2019} and to stabilization of different exotic structures,\cite{Makinen2019,VolovikZhang2020} including Bogoliubov Fermi surface\cite{Eltsov1908,Autti2020} and analogs of cosmic walls bounded by strings,\cite{Kibble1982}  see also review paper\cite{Kuang2021}.

The pinning may produce different routes in the transition from the polar phase to the $\beta$-phase under rotation. For example,
instead of the two HQVs, the elementary cell of the vortex lattice in the spin polarized polar phase may contain only singly quantized vortex in Eq.(\ref{SQV}). In this case it will continuously transform to the singly quantized vortex in the $\beta$ phase in Eq.(\ref{beta_vortex}).

 \begin{figure}[h]
 \includegraphics[width=0.7\columnwidth]{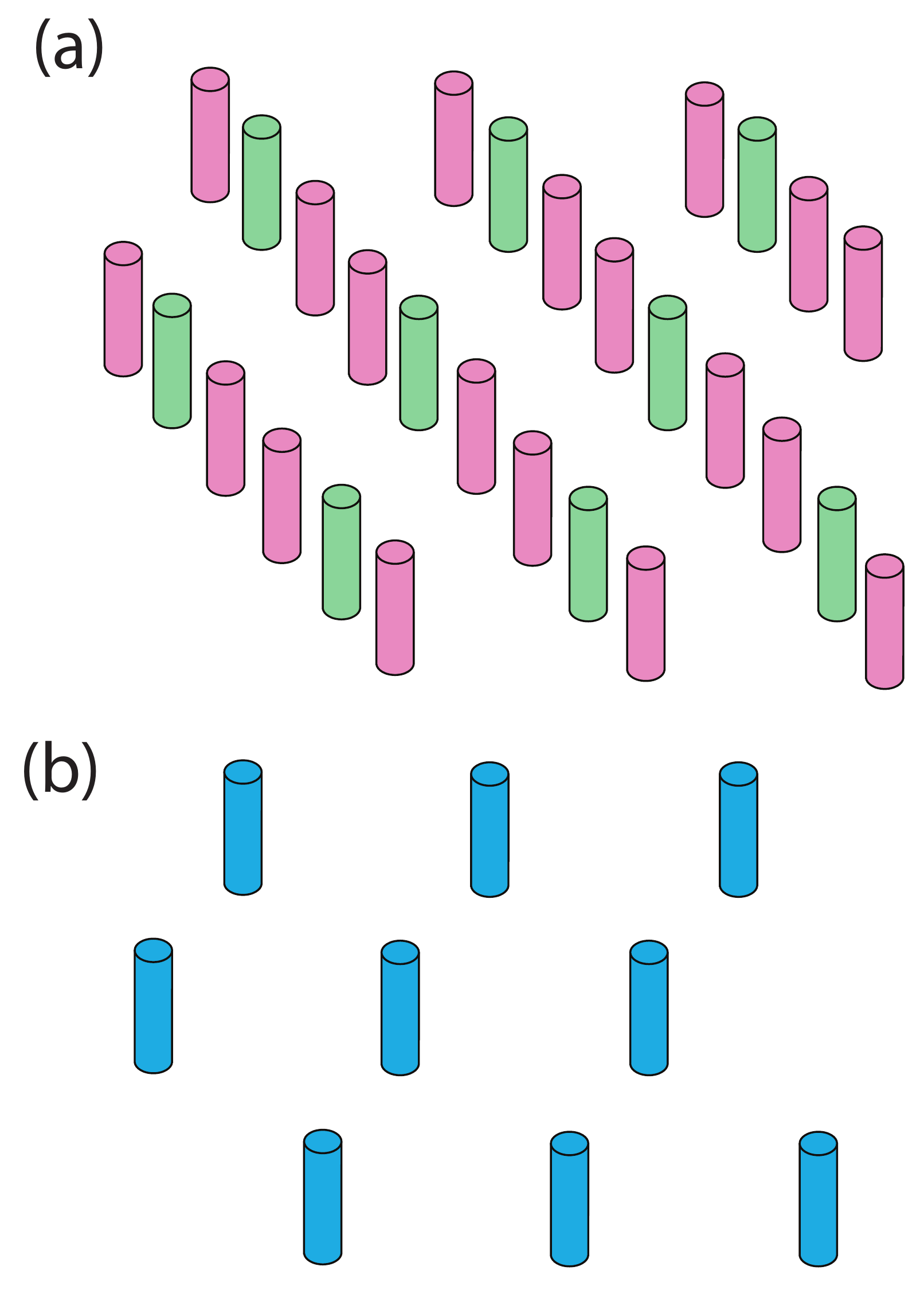}
 \caption{Illustration of  the rotating state in the  spin-polarized polar phase with three vortices in elementary cell.
(a) In this scenario, the elementary cell of spin-polarized polar phase contains three topological objects: two spin-down vortices in Eq.(\ref{down}) ({\it red}) and the spin vortex in Eq.(\ref{spin_vortex})  ({\it green}). With increasing field the intensity of the down-spin vortices decreases, and they finally disappear at the transition to the $\beta$-phase (b), while spin vortex  transforms to the single quantum vortex of the $\beta$ phase in Eq.(\ref{beta_vortex})  ({\it blue}).
 }
 \label{spin_vortices}
\end{figure}

The more complicated scenario is when  the elementary cell contains three objects:  two spin-down vortices in Eq.(\ref{down}) and the spin vortex (SV), see Fig.\ref{spin_vortices}(a). This spin vortex ({\it green}) has the following structure of the order parameter:
\begin{eqnarray}
{\bf A}_i= \hat z_i 
\left(\Delta_{\uparrow \uparrow}(\hat{\bf x} + i \hat{\bf y})e^{i\phi}  + \Delta_{\downarrow \downarrow}(\hat{\bf x} - i  \hat{\bf y})e^{-i\phi} \right) \,.
\label{spin_vortex}
\end{eqnarray}
The total energy of three objects is proportional to $(\rho_{s\uparrow}  + 3\rho_{s\downarrow} - 2\rho_{\uparrow\downarrow})$.

Approaching the $\beta$-phase, two spin-down vortices ({\it red}) fade away, while the SV continuously transforms to the SQV in the spin-up superfluid $\beta$-phase in Eq.(\ref{beta_vortex}), see Fig.\ref{spin_vortices}(b).

\section{Conclusion}

We considered topological objects in the polar phase and their transformation in the increasing magnetic field, when the polar phase becomes spin-polarized and then transforms to the $\beta$ phase. There are several scenarios of the topological evolution in the rotating cryostat, which include transformations of spin vortices, singly quantized vortices and two types of half-quantum vortices with opposite spin polarization.

  {\bf Acknowledgements}. I thank V.V. Dmtriev for discussions. This work has been supported by the European Research Council (ERC) under the European Union's Horizon 2020 research and innovation programme (Grant Agreement No. 694248).

\end{document}